\documentclass[showpacs,prl,twocolumn,aps,superscriptaddress]{revtex4}
\usepackage{epsfig}
\usepackage{graphicx}
\usepackage{amsmath}

\usepackage[colorlinks=true, pdfstartview=FitV, linkcolor=red, citecolor=blue, urlcolor=blue]{hyperref}

\begin{document}

\title{Decisive test of color coherence in proton-nucleus collisions at the LHC}

\author{Adam Bzdak}
\email{ABzdak@bnl.gov}
\affiliation{RIKEN BNL Research Center, Brookhaven National Laboratory, Upton, NY 11973,
USA}

\author{Vladimir Skokov}
\email{VSkokov@quark.phy.bnl.gov}
\affiliation{
Department of Physics, Brookhaven National Laboratory, Upton, NY
11973, USA; \\ 
Department of Physics, Western Michigan University, Kalamazoo, MI 49008, USA
}

\date{\today }

\begin{abstract}
Proton-nucleus collisions (p+A) at LHC energies provide a rigorous test 
of the Color Glass Condensate (CGC), a  model proposed  to describe 
the high energy limit of Quantum Chromodynamics. 
In the CGC the average multiplicity of charged particles at midrapidity in p+A collisions depends logarithmically
on the number of participants, $N_{\rm part}$. 
In contrast, the Wounded Nucleon Model (WNM) of independent nucleon-nucleon scatterings,
verified at RHIC energies, predicts that  
multiplicity in p+A depends linearly on $N_{\rm part}$.
We argue that
the dependence of mean multiplicity on $N_{\rm part}$ in p+A collisions at LHC energies
can single out a model of particle production,  
thus offering a stringent test of the CGC and the WNM. Based on this observation we 
propose a novel experimental test of color coherence in p+A collisions. 
\end{abstract}

\maketitle

Understanding the properties of strongly interacting matter at extreme conditions is 
the ultimate goal of
heavy-ion experiments at Relativistic Heavy-Ion Collider (RHIC) and the Large Hadron Collider (LHC).
The theoretical description of  nucleus-nucleus (A+A) collisions is
complicated
because of the importance of the non-perturbative nature of the strong forces at long distances, 
and the fact that one has to tackle a many body problem without obvious reduction to effective degrees of freedom. 
The last decade of collecting and analysing experimental data shows that a phenomenological application 
of  hydrodynamics, an effective theory of
long wavelength excitations in strongly-coupled systems, can successfully describe a large body of  
experimental data~\cite{Florkowski:book}, including momentum spectra  
of produced particles and two-particle correlations. However, from the theoretical perspective the application of 
hydrodynamics is not based on Quantum Chromodynamics (QCD) because many open issues are unresolved including the
non-equilibrium early stage of heavy-ion collisions, thermalization, and many others. These problems demand 
development of the 
non-perturbative methods to QCD, which are not available at present. Many models, including a holographic approach to QCD-like theories
(see e.g. Ref.~\cite{DeWolfe:2013cua}) 
and the Color Glass Condensate (CGC) (see e.g. Ref.~\cite{Gelis:2012ri})
aspire to describe various stages of a heavy-ion collision. However, a commonly 
accepted and complete picture is still lacking.  

The complexity of heavy-ion collisions is reduced considerably in the case of proton-proton (p+p) and 
proton-nucleus (p+A) collisions owing to the expected dominance of the initial state effects \cite{DuVe}, 
however recent studies of p+A collisions \cite{Bozek:2011if,BoBro,Bzdak:2013zma,Bzdak:2013lva} 
put this into question. The initial state effects are poorly studied in A+A collisions 
with their strong final state interactions.
Recent measurements at the LHC have provided 
new constraints on the models of particle production in p+p and p+A collisions, 
which can be used for further understanding of A+A collisions. Especially, 
disentangling the initial and final state effects is of primary importance 
for many studies of QCD properties in A+A collisions, as the search for the QCD Critical Point, 
CP violation and extracting 
transport properties of quark-gluon plasma.

One of the most important questions 
in p+A physics is whether the initial state of colliding nuclei behave as a superposition of its constituents 
or rather as a coherent gluon field predicted in the Color Glass Condensate. 
The CGC provides a framework to study particle production and scattering in QCD at high energies (small $x$). 
The key ingredients of the CGC are weak coupling, $\alpha_s\ll 1$;  large gluon density $\propto 1/\alpha_s$ corresponding to strong classical fields; nonlinear effects such as recombination and scattering of gluons characterized by single dynamically generated scale, $Q_s$,
the so-called saturation momentum, a typical transverse momentum below which the field modes have 
large occupation numbers $\propto 1/\alpha_s$.

At present, there is no direct experimental evidence in favor of the CGC. In this Letter, we provide 
a powerful yet simple method to disentangle the mechanism of particle production in p+A collisions
\footnote{It should be
noted that A+A collisions are not very effective in discriminating
between different models of particle production, see e.g., \cite{Kharzeev:2000ph}.}.
We show that in the CGC the mean number of produced particles at midrapidity depends logarithmically on 
the number of wounded nucleons, in a striking contrast to an expectation of 
a linear dependence on $N_{\rm part}$. In experiments $N_{\rm part}$ 
is difficult to measure in a model-independent way and we also propose an 
alternative probe based on the above observation, which can 
be straightforwardly measured at the LHC.

\begin{figure}[t]
\centerline{\includegraphics[width=0.38\textwidth]{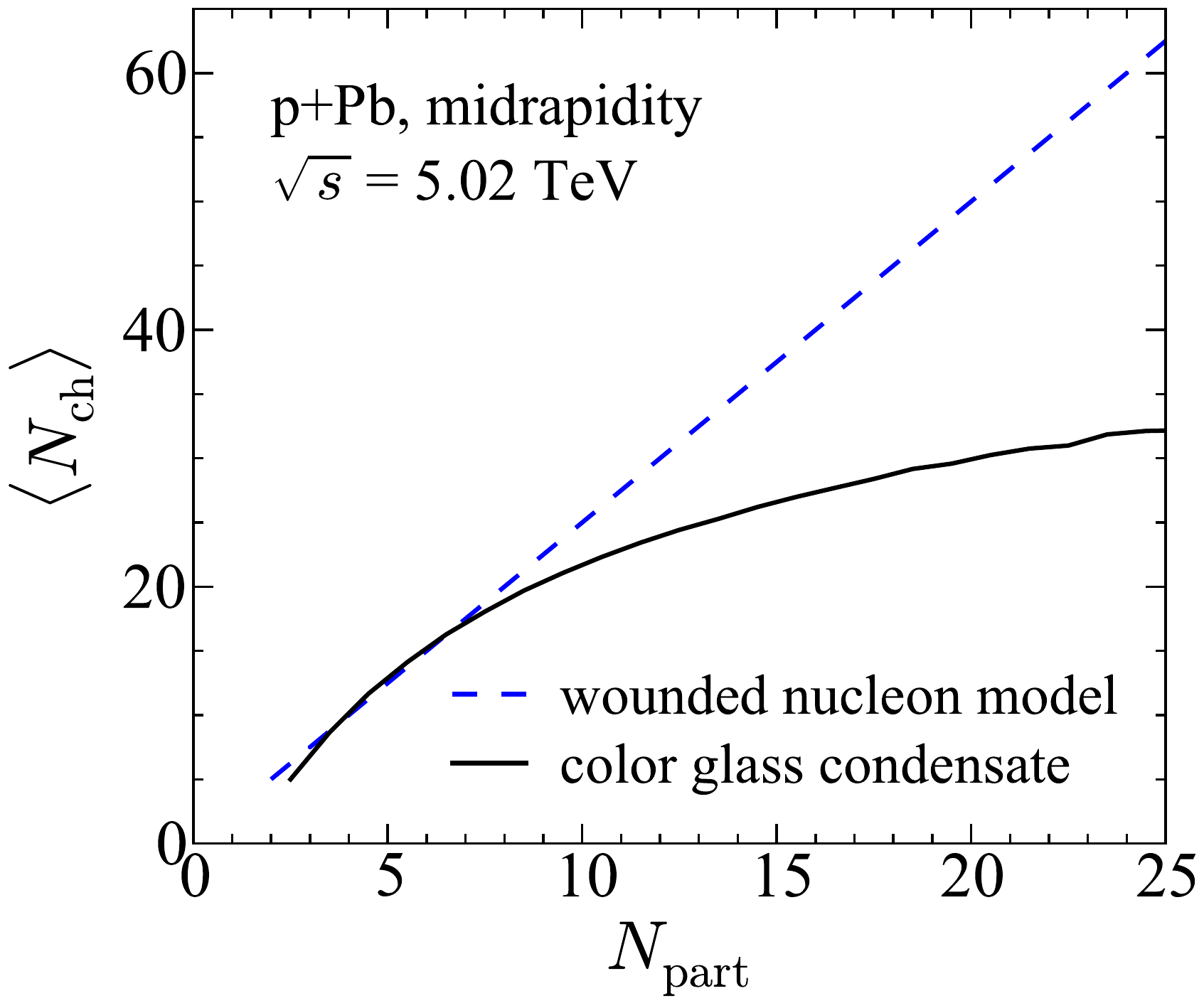}}
\caption{
The average number of particles at midrapidity, $\eta=0$, in the Color Glass Condensate (CGC) 
and the Wounded Nucleon model (WNM) in p+Pb collisions at $\sqrt{s}=5.02$ TeV. In the CGC,
the mean multiplicity depends logarithmically on  the number of participants. In the WNM, 
average multiplicity is a linear function of $N_{\rm part}$. In central p+A collisions,
corresponding roughly to $N_{\rm part}>18$, there is approximately a factor of $2$ 
difference between the two models. The CGC curve is computed numerically in the rcBK model~\cite{ALbacete:2010ad,Albacete:2012xq}. 
}
\label{fig:1}
\end{figure}

In the Wounded Nucleon Model (WNM)~\cite{Bialas:1976ed},
the average multiplicity in p+A 
collisions depends linearly on the number of participants
\begin{equation}
\langle N_{\rm ch}^{\rm pA} \rangle = \frac{\langle n_{\rm ch}^{\rm pp} \rangle}{2} N_{\mathrm{part}},
\label{wnm}
\end{equation}
where $\langle N_{\rm ch}^{\rm pA} \rangle$ and $\langle n_{\rm ch}^{\rm pp} \rangle$ are the 
average numbers of produced particles in
p+A and p+p collisions, respectively, and $N_{\mathrm{part}}$ is the number of wounded nucleons.
The interpretation of Eq.~\eqref{wnm} is the following. Each participant in a nucleus is struck exactly once 
and produces a certain number of particles, which, on average, is equal to the half of that 
in p+p collisions, $\langle n_{\rm ch}^{\rm pp} \rangle /2$. The projectile proton is assumed to 
yield the same number of particles 
independently of how many interactions it underwent~\footnote{This is justified for particles with
relatively small transverse momenta, because their formation time is longer than the time needed for 
a proton to pass through a nucleus. See Ref.~\cite{ABAB} for more details.}.
Equation (\ref{wnm}) was experimentally  verified for different energies,
ranging from 10 GeV~\cite{Elias:1979cp} to 200 GeV in d+Au collisions at RHIC ~\cite{Back:2004mr}. 
Particularly successful is the description of d+Au collisions: Eq. (\ref{wnm}) was confirmed for the total
number of produced particles~\cite{Back:2004mr} and their number at midrapidity~\cite{Bialas:2004su}. 
Moreover, with an extension of the WNM, the entire pseudorapidity distribution
at all measured centralities was described~\cite{Bialas:2004su,ABAB}.

\begin{figure}[t!]
\centerline{\includegraphics[width=0.38\textwidth]{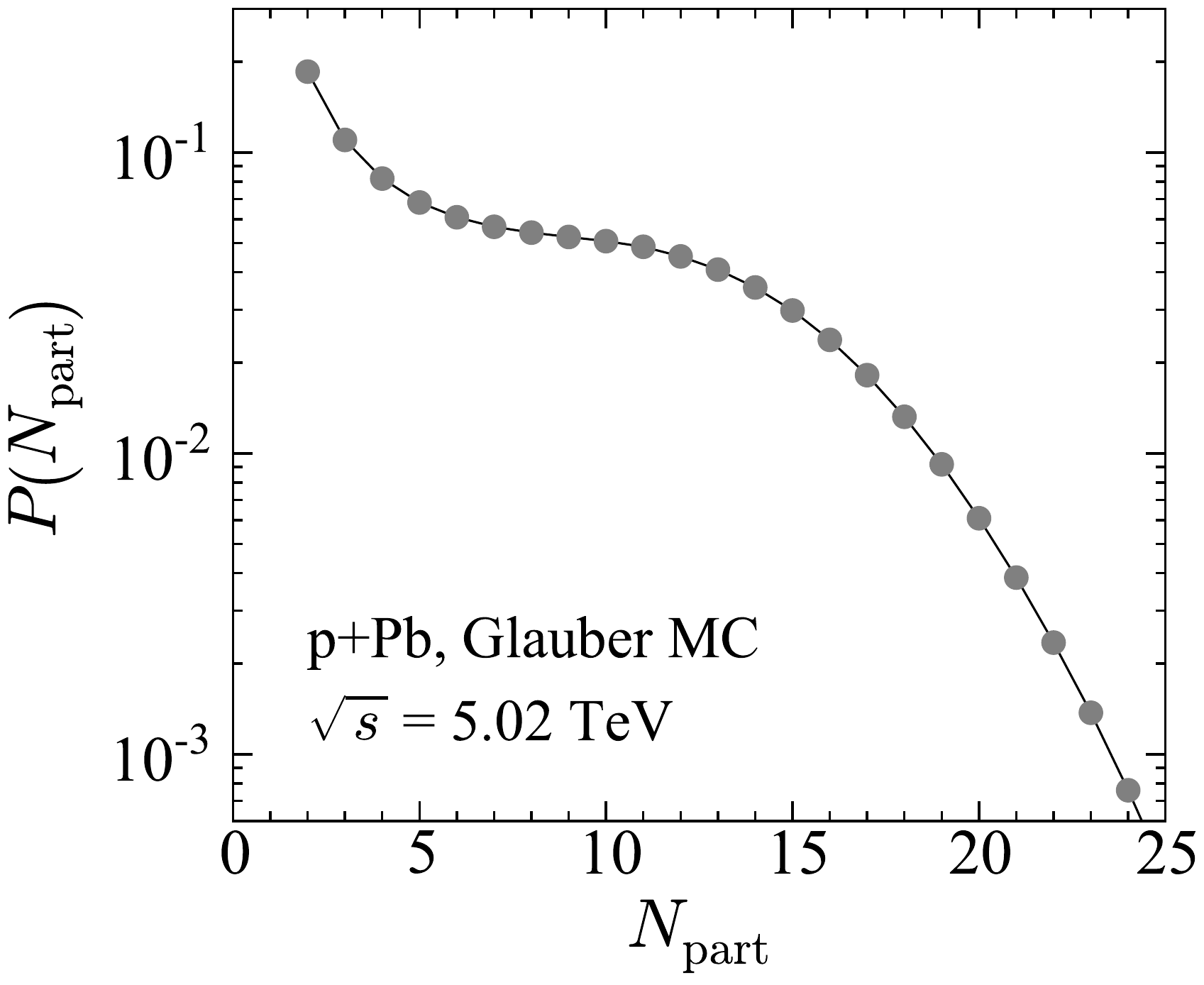}}
\caption{
The probability distribution of the number of wounded nucleons in p+Pb
collisions at $\sqrt{s}=5.02$ TeV in the Glauber Monte-Carlo model.
}
\label{fig:2}
\end{figure}

It was formerly believed that 
multiplicity in p+A depends on the number of collisions in accordance with  
\begin{equation}
\langle N_{\rm ch}^{\rm pA} \rangle = \langle n_{\rm ch}^{\rm pp} \rangle N_{\mathrm{coll}} = 
\langle n_{\rm ch}^{\rm pp} \rangle (N_{\mathrm{part}}-1).
\label{coll}
\end{equation}
In this scenario, each nucleon-nucleon collision in a nucleus produces on average 
$\langle n_{\rm ch}^{\rm pp} \rangle$ particles. 
In other words, a projectile proton creates new particles in every collision. Both Eqs. (\ref{wnm}) and (\ref{coll}) 
are characterized by a linear dependence on the number of participants, although in the former case 
particle abundance is significantly lower.  Equation (\ref{coll}) is expected to hold for the particles in 
the high transverse momentum region, $k_\perp\gg Q_s^{\rm (A)}$~\footnote{Here $Q^{\rm (A)}_s$ 
is the saturation momentum of a nucleus.},
where jets are produced in each nucleon-nucleon collision; while Eq.~(\ref{wnm}) 
is plausibly applicable 
for particles with transverse momenta in the non-perturbative region. 

\begin{figure*}[t]
\centerline{\includegraphics[width=0.75\textwidth]{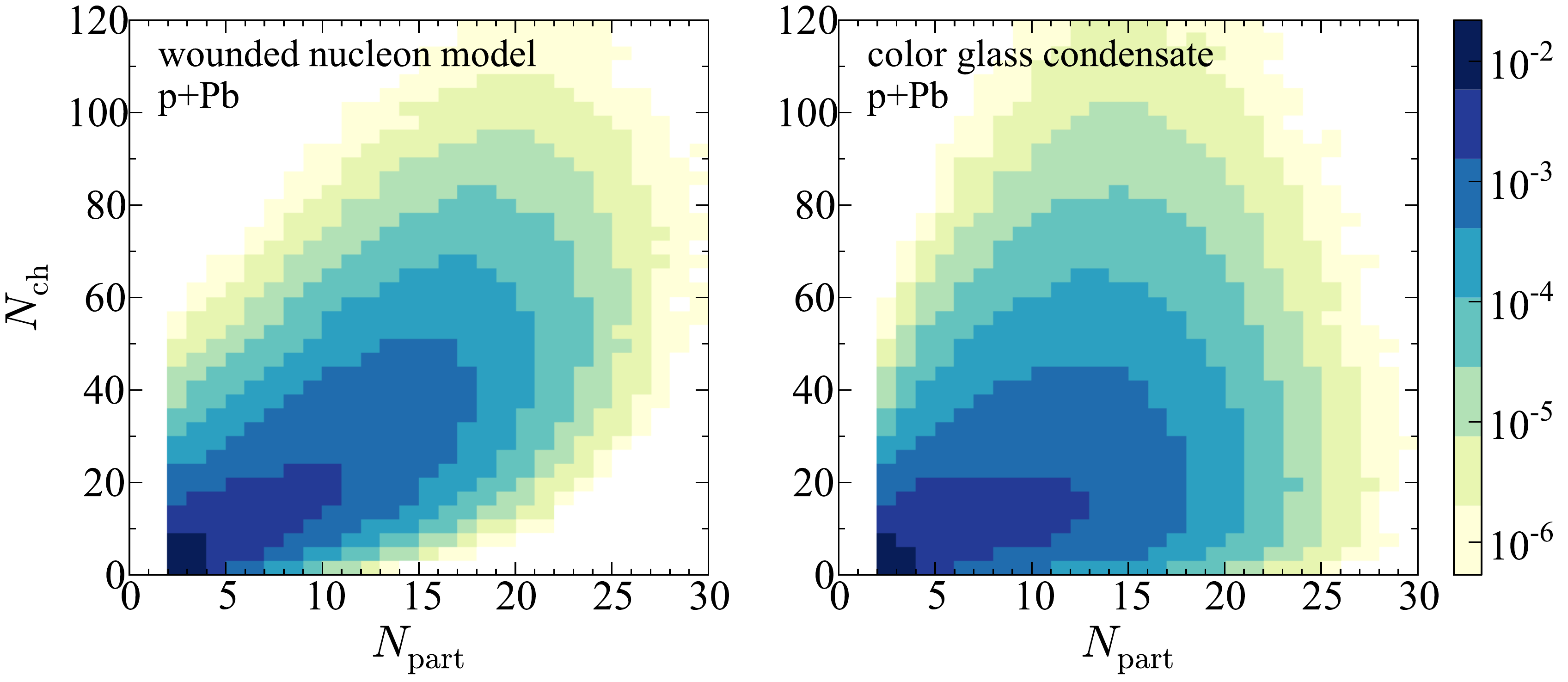}}
\caption{
The probability distribution of the number of participants, $N_{\rm part}$, and the number of 
produced particles at midrapidity, $N_{\rm ch}$, in the Wounded Nucleon Model (left) 
and the Color Glass Condensate (right) in p+Pb collisions at $\sqrt{s}=5.02$. 
For $N_{\rm part}>10$, the CGC has a significantly larger fluctuations of $N_{\rm ch}$ at a given 
number of $N_{\rm part}$.
}
\label{fig:3}
\end{figure*}  

The CGC result in the non-perturbative regime is different from Eq.~(\ref{wnm}). On very general grounds
it is predicted that at midrapidity
\footnote{
An analogous dependence was derived for A+A collisions in Ref.~\cite{Kharzeev:2000ph}:
$\langle N_{\rm ch} \rangle \sim   N_{\rm part} \ln N_{\rm part}$. }
\begin{equation}
\langle N_{\rm ch}^{\rm pA} \rangle \sim \ln(N_{\rm part}). 
\label{cgc}
\end{equation}
We will not dwell on the technical aspects of the derivation of Eq.~\eqref{cgc}, but pass on at once to 
its intuitive understanding.
The dependence~\eqref{cgc} can be obtained from the integrated gluon transverse momentum spectrum found in 
Ref.~\cite{Kovchegov:1998bi,Dumitru:2001ux}. 
This spectrum might be understood from the argumentation which follows. 
First of all, assuming validity of the CGC framework,
we implicitly accept that $Q_s^{\rm (p) }$, $Q_s^{\rm (A)}\gg  \Lambda_{\rm QCD}$. 
For the sake of argument, we also consider that the saturation momentum of a proton 
is much smaller then that of a nucleus, 
$Q_s^{\rm (p)}  \ll Q_s^{\rm (A)}$. In this case, the gluon spectrum 
can be obtained analytically under reasonable assumptions. 
Gluons at large transverse momentum, $k_\perp \gg  Q_s^{\rm (A)} $, are affected
by neither the field of the proton nor the nucleus, and consequently are produced according
to the ordinary perturbation theory of QCD with the characteristic $~1/k_\perp^4$ dependence.  
The low $k_\perp$ gluons, $\Lambda_{\rm QCD}<k_\perp<Q_s^{\rm(p)}$, are in the nonlinear 
regime with respect to both the fields of the proton and nucleus.  
The problem of gluon production in this domain is only tractable numerically.
Here, the gluon distribution is expected to be independent of $k_\perp$ 
modulo logarithmic corrections. 
In the intermediate region, $Q_s^{\rm(p)} < k_\perp < Q_s^{\rm (A)}$, the proton's gluon field 
can be treated perturbatively, while that of the nucleus is in the saturation regime and must be 
accounted for classically. 
The gluon production can effectively  be regarded as bremsstrahlung of large-$x$ partons 
in the classical gluon field of the nucleus. Thus, the transverse momentum dependence is modified 
from the perturbative one to
the one inherently characteristic for bremsstrahlung $\sim1/k^2_\perp$ (see also Ref.~\cite{Blaizot:2004wu}
for the numerical confirmation of this dependence). 

The gluon momentum distribution at midrapidity is dominated by this intermediate region 
$Q_s^{\rm(p)} < k_\perp < Q_s^{\rm (A)}$ with $d^2N/d^2k_\perp \sim 1/k_\perp^2 $ \footnote{The gluon 
fragmentation function is not expected to modify the conclusions of this article, which are drawn for 
the charged particle multiplicity.}. Integrating 
w.r.t. $k_\perp$, we obtain \footnote{It should be
noted that the different dependence of $\langle N_{\rm ch} \rangle$ on $N_{\rm part}$ in the WNM and CGC
leads to different values of $\langle N_{\rm part} \rangle$ in a given range of the
measured number of particles at midrapidity. Thus, to extract $\langle N_{\rm part} \rangle$
reliably one has to perform measurements near the nucleus rapidity,
where the number of particles depends linearly on $N_{\rm part}$ in both models.} 
\footnote{See also Ref.~\cite{Dumitru:2001ux}, where logarithmic corrections were taken into account.}
\begin{equation}
\langle N_{\rm ch} \rangle \sim \ln\left( \frac{Q_s^{\rm (A)}}{Q_s^{\rm (p)}} \right) \sim \ln (N_{\rm part}). 
\label{dndy}
\end{equation}
The second part of this equation, originates from the fact
that
$ (Q_s^{\rm (A)})^2$   is proportional to the number of participants
from the nucleus, $(Q_s^{\rm (A)})^2 \propto N_{\rm part} -1 \approx N_{\rm part}$. 
This discussion is deficient in several aspects. More rigorous approach is to incorporate quantum corrections 
by solving renormalization group equations, as the JIMWLK hierarchy~\cite{JIMWLK}, on top of classical field defined by  
the ab-initio first principle calculations of the nuclear wave function at  small $x$.  
The JIMWLK equations can be written in a closed form in the limit of large number of colors, the 
so called BK equation. The latter can be improved by 
taking into account the running coupling corrections. This approach was developed in 
Refs.~\cite{ALbacete:2010ad,Albacete:2012xq}	
and is referred to as rcBK in the literature. The initial conditions for the BK evolution are defined by
the global fits of the deep inelastic scattering data performed by the AAMQS collaboration \cite{Albacete:2010sy}. 
For more details of the
rcBK implementation and its predictions for p+A collisions we refer the reader to  
Refs.~\cite{ALbacete:2010ad,Albacete:2012xq}.

Figure~\ref{fig:1} illustrates the predictions of the CGC and the WNM for the average number of 
charged particles at midrapidity \footnote{It should be noted that the jet contribution can
slightly modify the CGC prediction. The projectile proton interacts
many times with the nucleons in a nucleus and an appropriate $N_{\rm coll}$ term
should be added (however, all nucleons in a nuclei are struck only once
and for them such term is not warranted). This would result in a
slightly faster increase of $\langle N_{\rm ch} \rangle$ with $N_{\rm part}$. We do not expect this
contribution to be larger than 10\% and we neglect it in this article.}, $\eta=0$, in p+Pb collisions at $\sqrt{s}=5.02$ TeV. 
In peripheral collisions with $N_{\rm part}=2$, corresponding to p+p, 
both models agree with the CMS measurement~\cite{Khachatryan:2010us}. 
The rcBK results demonstrate expected logarithmic dependence on the number of participants. 
To assure that this is not an artefact of neglecting the higher order correlations in the 
JIMWLK  evolution, the IP-Glasma model~\cite{Schenke:2012wb}
was used, which circumvents the problem of 
solving JIMWLK by parametrization of the saturation scale via the IP-Sat~\cite{Kowalski:2003hm}
model fitted to the HERA data. The IP-Glasma model reproduced the rcBK result with 
high precision~\footnote{We thank Bjoern Schenke for providing us with the IP-Glasma results.}.
We also verified Eq.~\eqref{cgc} in the KLN model \cite{Dumitru:2011wq}. 
This agreement of different implementations of the CGC suggests that Eq.~\eqref{cgc} is a universal 
property of the CGC.

As studied in Refs.~\cite{Kharzeev:2002ei,Bialas:2004su} both the WNM and CGC can
describe the data on d+Au collisions at RHIC. As seen
from Fig.~\ref{fig:1}, the difference between the two models is
noticeable in the region with high number of participants,
$N_{\rm part} > 12$. We argue that this region is accessible at the
LHC. Indeed, as demonstrated in  Fig.~\ref{fig:2}, 
the probability distribution of $N_{\rm part}$ 
in p+Pb collisions is high even for $N_{\rm part}=20$. This result is based on the
standard Glauber Monte-Carlo model with the Gaussian inelastic
nucleon-nucleon profile \cite{ABAB,Rybczynski:2011wv} and the inelastic cross section 
$\sigma = 70$ mb. 

It is a formidable problem to assess, in a model-independent way, the number of wounded nucleons. 
Consequently, the observation presented in Fig.~\ref{fig:1} may be challenging to test experimentally. 
Based on Fig.~\ref{fig:1}, one may speculate, that the WNM and the CGC should result in   
very different multiplicity distributions at high numbers of produced particles, $N_{\rm ch}$.
This is because
events with high $N_{\rm ch}$ are correlated
with those having high number of participants, where the color coherence effects are best visible. 
However, this is not supported by direct calculations. 
As shown in Fig.~\ref{fig:3}, for a large number of participants, $N_{\rm part}>10$, 
fluctuations of the number of produced particles at a given $N_{\rm part}$ 
are significantly larger in the CGC than in the WNM \footnote{In both models we generated $5\times 10^{6}$ events.}. 
Consequently the projected multiplicity distributions of
the model predictions in Fig.~\ref{fig:3} are approximately the same within the
uncertainties of the models \footnote{In the WNM each participant populates particles according 
to the negative binomial distribution. We use $\langle n_{\rm ch}^{\rm pp} \rangle = 5$ and $k_{\rm pp} = 1.1$.}.

\begin{figure}[t]
\centerline{\includegraphics[width=0.38\textwidth]{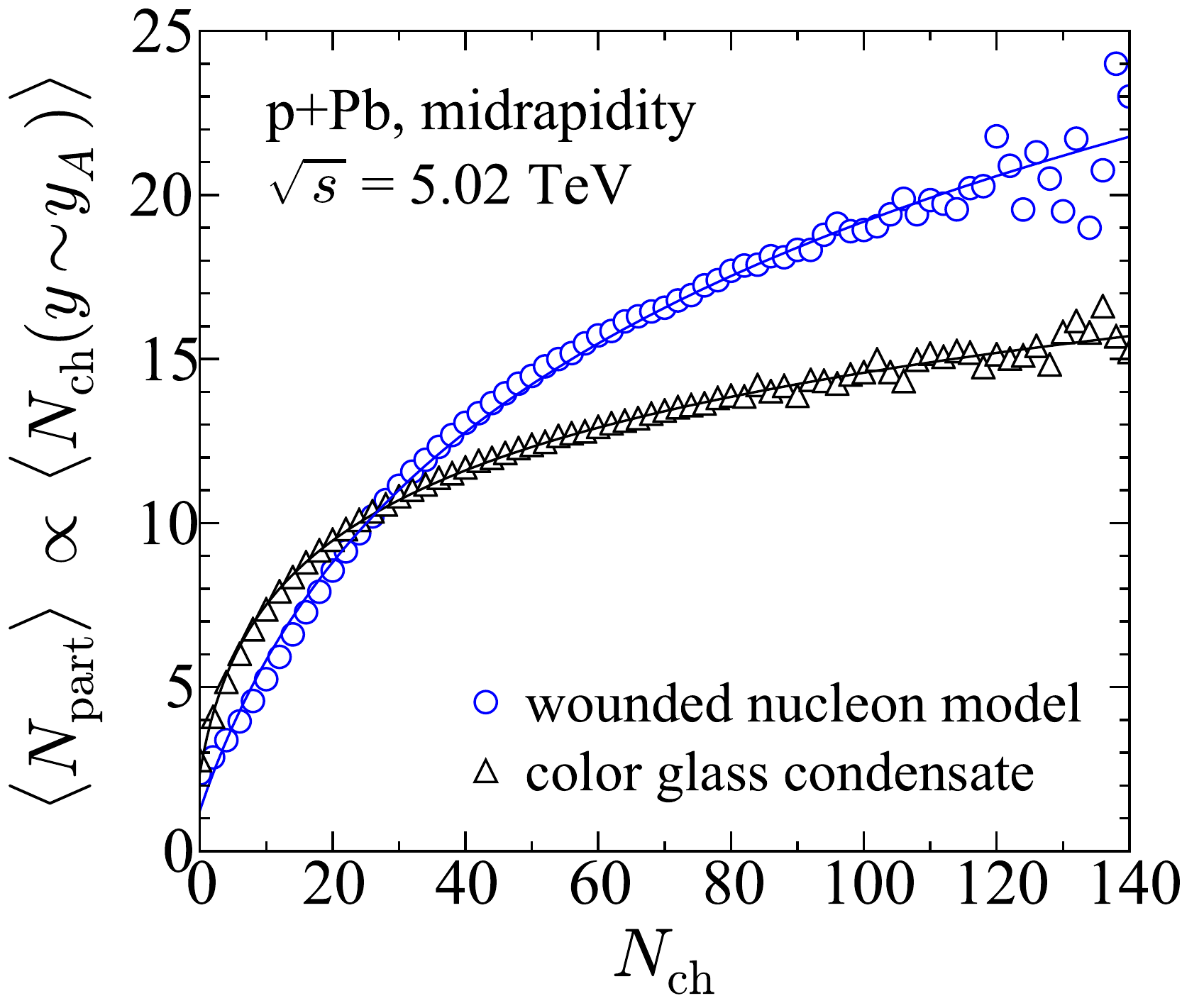}}
\caption{
The average number of participants, $\langle N_{\rm part} \rangle$, at a given number of 
produced charged particles at midrapidity, $N_{\rm ch}$, in the Wounded Nucleon Model and the Color Glass Condensate 
in p+Pb collisions at $\sqrt{s}=5.02$. 
The average number of participants can be replaced by the mean number of particles in a (broad) vicinity of a nucleus fragmentation region, $\langle N_{\rm ch}(y \sim y_A) \rangle$, where $y_A$ 
is a nucleus rapidity (see the text for more details). Alternatively $\langle N_{\rm part} \rangle$ could be 
estimated via the number of high $k_{\perp}$ particles (pions, photons etc.) that are known to scale with the number of collisions, which is equivalent to $N_{\rm part}$ in p+A.
}
\label{fig:4}
\end{figure}

In this Letter we propose a novel approach to discriminate 
between the CGC and the WNM. 
In Fig.~\ref{fig:4} we present the relation between the average number 
of participants at a given number of 
produced charged particles at midrapidity. 
The CGC and the WNM lead to very different relations especially 
for low and large number of produced particles. For example, 
at $N_{\rm ch} = 140$ the mean number of participants differs approximately by $50$\%, 
being larger in the WNM than in the CGC. At $N_{\rm ch} = 10$ the
situation is opposite with more participants, on average, in the CGC than the WNM. 
The latter can be deduced from Figs.~\ref{fig:1} and ~\ref{fig:3}.
Indeed, in the CGC,
events with a broad range of $N_{\rm part}$ can deliver a small
number of particles, wherein in the WNM only events with small 
$N_{\rm part}$ can yield small $N_{\rm ch}$. At large $N_{\rm ch}$, strong fluctuations in
the CGC can lead to a large number of produced particles even if $N_{\rm part}$ is not particularly large. 

One could argue that $\langle  N_{\rm part} \rangle$ is as difficult to measure as $N_{\rm part}$,
which is, however, not the case. The mean number of participants at a given ${N_{\rm ch}}$ at zero (pseudo)rapidity 
can be extracted in several ways. Below, we list them in descending order 
of theoretical reliability and experimental accessibility.     
\begin{enumerate}

\item Number of charged particles in a vicinity of the nucleus fragmentation region, 
$\left. dN/d\eta \right|_{\eta \sim \eta_A }$.
For sufficiently high (pseudo)rapidity it is proportional to $(N_{\rm part}-1)$ modulo 
small corrections. This observable can be further improved by considering the difference 
$(\left. dN/d\eta \right|_{\eta = \eta^* } - \left. dN/d\eta \right|_{\eta = -\eta^*})$, 
in which central rapidity effects cancel out. 
We checked that the ratio of this combination to $\langle N_{\rm part} \rangle$ is independent of $N_{\rm ch}$ 
in the rcBK model already at the moderate values of $\eta^*\geq2$. 

\item  It is known that at large $k_{\perp}$ particles scale with the number of collisions,
$N_{\rm coll} = N_{\rm part} - 1$ in p+A. This is the case in the CGC, where 
$R_{\rm pPb}=1$~\cite{Albacete:2012xq}. 
Thus the average number of participants in Fig.~\ref{fig:4} can be replaced by
the number of high $k_{\perp}$ particles.
 
\item The observable proposed in 2), can also be applied to high $k_\perp$ photons. 
\end{enumerate}

In conclusion, we argued that p+A collisions at the LHC have potential to produce the 
evidence for the Color Glass Condensate. 
The CGC predicts a logarithmic dependence of the midrapidity average number
of produced particles on the number of participants (wounded nucleons). This is 
a direct manifestation of the color coherence effects present in the CGC. 
As we demonstrated, in central p+Pb collisions the mean number of particles at midrapidity in the CGC
is expected to be smaller approximately by a factor of $2$ compared to predictions of the Wounded Nucleon 
Model, where the mean number of particles depends linearly on $N_{\rm part}$. 
Similar dependence  is  expected to be present in all models based on the Glauber
initial conditions disregarding color coherence effects. Based on this observation we proposed
a novel and straightforward  measurement, which can be carried out at the LHC.

We thank Adrian Dumitru, Larry McLerran and Al Mueller  for valuable comments and in-depth discussions. 
Comments by Robert Pisarski, Bjoern Schenke, Paul Sorensen and Raju Venugopalan 
are acknowledged.
A.B. is supported through the RIKEN-BNL Research Center. 
V.S. is supported 
by the U.S. Department of Energy under contract \#DE-AC02-98CH10886.

{\it Note added. ---}
After submission of our paper to 
Physical Review Letters, preliminary
data were presented by the ATLAS
Collaboration on p+Pb collisions \cite{atlas-prel}. 
The dependence of the number of charged 
particles on the number of participants 
appears to be close to linear, suggesting 
that the WNM is applicable at LHC energies.
This conclusion is sensitive, however, to how 
the number of participants is extracted in p+Pb collisions.

\end{document}